\documentstyle[epsfig,aps,prl]{revtex}

\newcommand{\smfrac}[2]{\mbox{$\frac{#1}{#2}$}}

\title{Generalized Drude model: Unification of ballistic and
diffusive electron transport
}
\author{
R.\ Lipperheide, T.\ Weis, and U.\ Wille}
\address{
Abteilung Theoretische Physik, Hahn-Meitner-Institut Berlin,\\
Glienicker Str.\ 100, D-14109 Berlin, Germany}
\begin{document}
\date{\today}
\maketitle

\begin{abstract}
For electron transport in parallel-plane semiconducting structures, a model is
developed that unifies ballistic and diffusive transport and thus generalizes
the Drude model.  The unified model is valid for arbitrary magnitude of the
mean free path and arbitrary shape of the conduction band edge profile.
Universal formulas are obtained for the current-voltage characteristic in the
nondegenerate case and for the zero-bias conductance in the degenerate case,
which describe in a transparent manner the interplay of ballistic and diffusive
transport.  The semiclassical approach is adopted, but quantum corrections
allowing for tunneling are included.  Examples are considered, in particular
the case of chains of grains in polycrystalline or microcrystalline
semiconductors with grain size comparable to, or smaller than, the mean free
path.  Substantial deviations of the results of the unified model from those of
the ballistic thermionic-emission model and of the drift-diffusion model are
found.  The formulation of the model is one-dimensional, but it is argued that
its results should not differ substantially from those of a fully
three-dimensional treatment.  \\[0.3cm]

\noindent
PACS number(s): 05.60.-k, 05.60.Cd, 72.10.-d, 72.20.-i
\end{abstract}

\section{Introduction}

Electron transport in semiconducting structures is {\em ballistic} if the mean
free path is much larger than the characteristic dimensions of the sample, and
it is {\em diffusive} if the mean free path is much smaller than these.  In the
first case, there is no impurity or lattice scattering, and the current is
determined by the ballistic motion in the electric field \cite{bet42,sze81}; in
the second case, scattering predominates and is described within the
drift-diffusion scheme \cite{sch38,mot38}.  In many cases of physical
relevance, however, the mean free path is neither large nor small compared with
the characteristic dimensions of the sample.  Thus, formulas for the
current-voltage characteristic have appeared in the literature which combine
features of the two limiting types of transport mechanism for particular
conduction band edge profiles, e.g., for a single barrier
\cite{cro66,mcg83,sim87,eva91,mck61,lug81,pri98}.

In the present paper, we consider electron transport in parallel-plane
semiconducting structures, i.e., structures whose parameters vary in one
direction only.  We develop a one-dimensional transport model which is valid
for any magnitude of the mean free path and any form of the band edge profile,
and thus unifies the ballistic and diffusive transport mechanisms.  It is based
on the idea that the electrons move ballistically in the electric field over
intervals with average length equal to a universal mean free path, after which
they are thermalized into a state of local equilibrium characterized by a
quasi-Fermi level (electrochemical potential).  The length of the sample is
made up of random configurations of such ballistic intervals.  Averaging over
these configurations results in a unified description of electron transport, in
which purely ballistic and purely diffusive transport appear as limiting cases.
We work within the semiclassical approach \cite{ash76}, which allows a concise
and transparent formulation.  However, quantum tunneling (``thermionic field
emission'' \cite{dol54,str62}) is taken into account in WKB approximation.

The description of transport in terms of ballistic motion over intervals of the
average length of the mean free path with thermalization at the end is, of
course, also the basis of the Drude model \cite{ash76,dru00,sap95} and of the
relaxation-time approximation of the Boltzmann equation.  There, however, the
further development makes use of the assumption that the mean free path is
small compared to the characteristic dimensions of the sample, leading to a
diffusive description of the transport.  In contrast to this, such an
assumption is {\em not} made in the present work, and the magnitude of the mean
free path relative to the sample dimensions determines the relative importance
of the ballistic and diffusive transport mechanisms.  The unified description
thus is a generalization of the Drude model; it is particularly relevant to
polycrystalline and microcrystalline materials (in the following
indiscriminately referred to as ``polycrystalline materials'') when the grain
size is comparable to, or smaller than, the mean free path.  In this case, the
grains must not be considered separately and the sample must be treated as a
whole.

Our approach is the semi-phenomenological one commonly used in the description
of transport in semiconductor devices (cf., e.g., Ref.\ \cite{sze81}).  Its
prominent and useful feature is that its results are obtained in closed form,
which allows one to analyze the physics of the transport process in a
transparent way.  We have been able to formulate our model in one dimension.
For transport across parallel-plane structures as considered in the present
work, a fully three-dimensional treatment of the thermalization process is
expected not to change the resulting formulas in a substantial way (cf.\
Sec.\ III.D below for a more detailed discussion of this point).

In the following section, we introduce the basic assumptions of the unified
model and explain the procedure of averaging over the random configurations of
ballistic intervals across which the electrons travel without scattering.  In
Sec.\ III, we present our principal result:  a universal formula for the
current-voltage characteristic reflecting the interplay of ballistic and
diffusive transport.  This formula generalizes previous expressions proposed
for a single barrier.  Its distinctive new feature is the appearance of a shape
term that depends on the detailed structure of the band edge profile and also
explicitly on the mean free path.  This term is essential, e.g., for the
description of transport in polycrystalline materials.  Section IV deals with
two numerical examples.  First, we investigate chains of identical grains in
polycrystalline materials.  Here, the effect of the relative magnitudes of mean
free path and characteristic length of the sample, i.e., the relative
importance of the ballistic and diffusive mechanisms, is demonstrated
explicitly.  It is found that the (zero-bias) conductivity, defined as
conductance times sample length, generally depends upon the number of grains in
the sample.  Second, as an example of a degenerate system, we discuss the grain
barrier conductance of a highly doped polycrystalline material as a function of
temperature.  Section V contains a summary and some concluding remarks.

\section{The model}

\subsection{Basic formulation}

\begin{figure}
\epsfysize=5.0cm
\epsfbox[10 577 358 725]{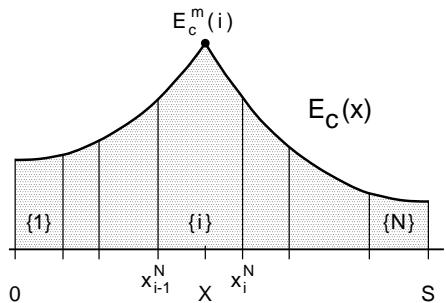}
\caption{
Averaging over the ballistic configurations: single peak of the band edge
profile $E_{\rm c}(x)$ at $x=X$.
}
\vspace{0.4cm}
\label{fig:1}
\end{figure}

The unified model is based on the following one-dimensional scheme (cf.\ Fig.\
\ref{fig:1}).  The length of the sample (extending from $x=0$ to $x=S$) is
covered by a chain of $N$ intervals $\{i\}$ with end points $x_{i-1}^{N}$ and
$x_{i}^{N}$, $i=1,...,N$ ($x_{i}^{N} > x_{i-1}^{N}; x_{0}^{N} = 0, x_{N}^{N} =
S$) across which the electrons move ballistically in the field of the
conduction band edge potential $E_{\rm c}(x)$.  There may be any number of
intervals in a chain, $N =1,..., \infty.$ At the end points of these
``ballistic intervals'' (in particular, at the fixed end points of the sample),
the electrons are equilibrated (thermalized).  These points of local
equilibrium are theoretical constructs which may be thought to be connected via
thin ideal leads (flat potentials) to large ideal reservoirs in equilibrium,
with chemical potentials equal to the quasi-Fermi level at the points
\cite{bar89}.  The current flow in a ballistic interval is assumed to result
from the injection of electrons at the end points and their transmission
through the interval, in line with Landauer's view of conduction as a
transmission phenomenon \cite{imr99,lan87,lan89,sto88}.

The current in the ballistic interval $\{i\}$ is given by
\begin{equation}
j(i) = -e \int_{0}^{\infty} d \epsilon \, \left[ T_i^{\rm L}(\epsilon + E_{\rm
c}(x_{i-1}))f(x_{i-1}^{N}; \epsilon)
- T_i^{\rm R}(\epsilon + E_{\rm c}(x_{i})) f(x_{i}^{N};
\epsilon) \right] \; , \hspace{1cm} i= 1,...,N \; ,
\label{eq:0}
\end{equation}
where $f(x;\epsilon)$ is the phase space density of the electrons at position
$x$ with kinetic energy $\epsilon$ of the motion in the $x$-direction, and
$T_i^{\rm L}(E)$ is the (classical or quantal) probability for ballistic
transmission at total energy $E$ from $x_{i-1}^{N}$ to $x_{i}^{N}$ [reversely
for $T_i^{\rm R}(E)$].  Owing to time reversal invariance we have $T_i^{\rm
L}(E)=T_i^{\rm R}(E)[=T_i(E)]$.  For Boltzmann statistics (nondegenerate
regime), we have $f(x;\epsilon) = (4\pi m^*/\beta h^3) \exp\{-\beta[\epsilon +
E_{\rm c}(x) - E_{\rm F}(x)]\}$, and obtain from Eq.\ (\ref{eq:0})
\begin{equation}
j(i) = -e v_{\rm e} N_{\rm c} \, {\cal T}(i)
\left[e^{\beta E_{\rm F}(x_{i-1}^{N})}-e^{\beta E_{\rm
F}(x_{i}^{N})}\right] \; , \hspace{1cm} i= 1,...,N \; ;
\label{eq:1aa}
\end{equation}
here, $v_{\rm e} = (2 \pi m^{*} \beta)^{-1/2}$ is the emission velocity, the
factor $N_{\rm c} = 2 (2\pi m^{*}/\beta h^{2})^{3/2}$ is the effective density
of states at the conduction band edge, $E_{\rm F}(x)$ is the quasi-Fermi level
at position $x$, and $\beta = 1/k_{\rm B}T$.  The factor ${\cal
T}(i)$ is the thermally averaged  probability for ballistic transmission across
the interval $\{i\}$,
\begin{equation}
{\cal T}(i) = \beta \int_{E_{\rm c}^{(i)}}^{\infty} dE \,
e^{-\beta E} \, T_{i}(E) \; ,
\label{eq:1aaa}
\end{equation}
where $E_{\rm c}^{(i)} = {\rm max} \{ E_{\rm c}(x_{i-1}), E_{\rm c}(x_{i}) \}$
(cf.\ Ref.\ \cite{duk69}, and Ref.\ \cite{yan93}, Sec.\ 2.1).

In the classical description, we have $T_{i}(E) = \Theta (E -  E^{\rm m}_{\rm c}(i))$,
where $E^{\rm m}_{\rm c}(i)$ is the maximum of the conduction band edge
profile $E_{\rm c}(x)$ between or at the points $x_{i-1}^{N}$ and $x_{i}^{N}$.
Therefore, Eq.\ (\ref{eq:1aa}) becomes
\begin{equation}
j(i) = -e v_{\rm e} N_{\rm c} \, e^{-\beta E^{\rm m}_{\rm c}(i)}
\left[e^{\beta E_{\rm F}(x_{i-1}^{N})}-e^{\beta E_{\rm F}(x_{i}^{N})}\right] \;
, \hspace{1cm} i= 1,...,N \; .
\label{eq:1}
\end{equation}
In what follows, we adhere to the classical description since it allows the
greatest transparency, in particular as far as the averaging over ballistic
intervals is concerned. The full formulation including tunneling effects will
be presented in Sec.\ III.B.

In the stationary case, the current is independent of position, $j(i) = j = $
const., and we can write Eq.\ (\ref{eq:1}) in the form
\begin{equation}
e^{\beta E_{\rm F}(x_{i}^{N})} = e^{\beta E_{\rm F}(x_{i-1}^{N})} +
\frac{j}{ev_{\rm e}N_{\rm c}} \, e^{\beta E^{\rm m}_{\rm c}(i)} \; ,
\hspace{1.0cm} i = 1, ..., N \; .
\label{eq:2}
\end{equation}
Iterating this relation, we find for a configuration with $N$ ballistic intervals
\begin{equation} e^{\beta E_{\rm F}(S)} - e^{\beta E_{\rm F}(0)} =
\frac{j}{ev_{\rm e}N_{\rm c}} \sum_{i =1}^{N} e^{\beta E^{\rm m}_{\rm c}(i)} \;
.
\label{eq:3}
\end{equation}
The sum on the right-hand side must be averaged over all possible configurations
of the ballistic
intervals, i.e., over all positions of their end points $x^{N}_{i}$ $(i=
1,...,N-1)$ in the chain of $N$ intervals, where $N = 1,..., \infty$.
Introducing the absolute maximum $E_{\rm c}^{\rm m}$ of the band edge
profile $E_{\rm c}(x)$ in the interval $[0,S]$, we denote the product of
$\exp(-\beta E_{\rm c}^{\rm m})$ with the average of the sum by $\Xi$,
\begin{equation}
\Xi =  \left\langle \sum_{i =1}^{N} e^{- \beta [E^{\rm m}_{\rm c} -
E^{\rm m}_{\rm c}(i)]} \right\rangle_{\{N,x^N_{i}\}} \; . \label{eq:4}
\end{equation}
Setting $E_{\rm F}(0) - E_{\rm F}(S) = eV$, we then have
\begin{equation}
1 - e^{-\beta eV} = - \frac{j}{ev_{\rm e}N_{\rm c}} \, e^{\beta E_{\rm p}}
\, \Xi \; , \label{eq:5}
\end{equation}
where $E_{\rm p} = E_{\rm c}^{\rm m}- E_{\rm F}(0)$ is the overall barrier
height. The end points of the sample are connected to large ideal reservoirs
in equilibrium characterized by the Fermi levels $E_{\rm F}(0)$ and $E_{\rm
F}(S)$ whose difference determines the voltage bias $V$ \cite{lan89}.

\subsection{Averaging over the ballistic configurations}

\subsubsection{The distribution of the ballistic intervals}

The distribution of a set of $N$ ballistic intervals, i.e.,\ of the points of
equilibrium $x^N_i$, across the sample of length $S$ is determined in the
following way.  The probability of an electron to make a collision in the
interval $d\xi$ after having traversed a distance $\xi$ since its last
collision is given by $\exp(-\xi/l)(d\xi/l)$, where $l$ is the mean free path
\cite{ash76,sap95}.  Assuming (as in the Drude model) that an electron which
collides with an impurity is taken out of the ballistic current and can be
counted as equilibrized, we write the complete distribution of the
ballistic intervals in the form of an infinite-dimensional {\em diagonal
matrix} with elements labelled by the number $N$ of ballistic intervals, $N=1,
2,...$ :
\begin{equation}
d{\bf P} = \left\{ e^{-S/l}, ..., \left[ \, \prod_{i=1}^{N-1}\frac{dx^N_i}{l}\;
e^{-(x_i^N - x_{i-1}^N)/l} \; \theta(x^N_i - x^N_{i-1})\right]  e^{-(S -
x_{N-1}^N)/l} , ... \right\}  =  \{..., dP_{N}, ...  \} \; ;
\label{eq:6}
\end{equation}
here the exponents in the general term cancel out except for the term $(-S/l)$,
and we have (recalling $x_{0}^{N} = 0$)
\begin{equation}
dP_{1} = e^{-S/l} \; ; \hspace{0.5cm} dP_{N} = e^{-S/l}
\prod_{i=1}^{N-1}\frac{dx^N_i}{l} \; \theta(x^N_i - x^N_{i-1}) \; \; \; {\rm
for} \;\; \; N \geq 2 \; .
\label{eq:6a}
\end{equation}
This distribution is normalized to unity, since
\begin{equation}
{\rm Tr} \int d{\bf P} = \left(1 + \sum_{N=2}^{\infty}
\int_0^{S}\frac{dx^N_1}{l} \int_{x_{1}^{N}}^{S}\frac{dx^N_2}{l}...
\int_{x_{N-2}^{N}}^{S} \frac{dx^N_{N-1}}{l}\right) e^{-S/l} = 1 \; .
\label{eq:7}
\end{equation}
Here, the infinite sum over $N$ contains power terms which add up to the
exponential exp$(S/l)$; the evaluation of all quantities considered in the
following runs along similar lines.

The average sum (\ref{eq:4}) can be written as an expectation value of the form
\begin{equation}
\Xi =  \left\langle \sum_{i =1}^{N} e^{- \beta [E^{\rm m}_{\rm c} -
E^{\rm m}_{\rm c}(i)]} \right\rangle_{\{N,x^N_{i}\}}  =  \; \; {\rm Tr} \int
d{\bf P} \, {\bf \Sigma},
\label{eq:15}
\end{equation}
where we have introduced the infinite-dimensional diagonal matrix
\begin{equation}
{\bf \Sigma} = \left\{..., \sum_{i =1}^{N} e^{- \beta [E^{\rm m}_{\rm
c} - E^{\rm m}_{\rm c}(i)]}, ...  \right\} \; , \; \; N=1,2, \ldots \; \; .
\label{eq:16}
\end{equation}

\subsubsection{The average for a peak in $E_{\rm c}(x)$}

The case where the conduction band edge profile $E_{\rm c}(x)$ has a single
peak at the position $X$ somewhere along the sample is illustrated in Fig.\
\ref{fig:1}.  If a ballistic interval $\{i\}$ contains $X$, we have $E_{\rm
c}^{\rm m}(i) = E_{\rm c}(X)$; if it lies to the left [right] of $X$, we have
$E_{\rm c}^{\rm m}(i) = E_{\rm c}(x^N_i)$ $[ = E_{\rm c}(x^N_{i-1})]$.  Thus we
obtain for the average sum
\begin{eqnarray}
\Xi &=& \sum_{N=1}^{\infty} \int dP_{N} \sum_{i =1}^{N} e^{-\beta
E_{\rm c}^{\rm m}} \, \left[ e^{\beta E_{\rm c}(x^N_i)} \, \theta(X-x^N_i) +
e^{\beta E_{\rm c}(X)} \; \theta(x^N_i -X) \;
\theta(X-x^N_{i-1}) +  e^{\beta E_{\rm
c}(x^N_{i-1})} \, \theta(x^N_{i-1} -X) \right] \nonumber \\ &=&
1 + \widetilde{S}/l  \; ,
\label{eq:20}
\end{eqnarray}
where $E_{\rm c}^{\rm m} = E_{\rm c}(X)$ in the ``reduced sample length''
\begin{equation}
\widetilde{S} = \int_{0}^{S} dx \,  e^{- \beta [E_{\rm c}^{\rm m} -
E_{\rm c}(x)]}
\label{eq:27aa}
\end{equation}
satisfying $\widetilde {S} \leq S$. Special cases are $X = 0$ and $X = S$, when
the profile is monotonic.

It is seen that the average sum $\Xi$ for a profile containing a single maximum
at a position $X$ inside or at the end of the sample, is given by unity plus
the ratio of reduced sample length and mean free path.  By the definition of
the average sum $\Xi$, the unit term represents the contribution of
the ballistic transmission across the {\em highest} peak $E_{\rm c}^{\rm m}$ of
the profile $E_{\rm c}(x)$.

\subsubsection{The average for a valley in $E_{\rm c}(x)$}

\begin{figure}
\epsfysize=5.0cm
\epsfbox[40 641 257 750]{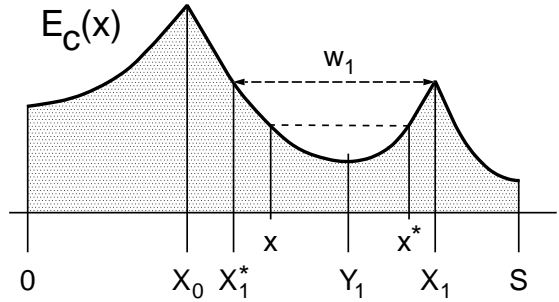}
\caption{
Averaging over the ballistic configurations:  single valley of the band edge
profile $E_{\rm c}(x)$.  For explanation, see text.
}
\vspace{0.4cm}
\label{fig:2}
\end{figure}

We consider a conduction band edge profile of the type shown in Fig.\
\ref{fig:2}. It contains two peaks at $X_{0}$ and $X_{1}$, respectively
(without loss of generality, the left peak is assumed to be the higher one),
and in between a valley with minimum at $Y_{1}$. The average sum over the
ballistic intervals {\em enclosed} by the peaks $(X_{0} \leq x_{i-1}^{N}$;
$x_{i}^{N} \leq X_{1})$ is given by
\begin{eqnarray}
\Xi &=& \sum_{N=1}^{\infty} \int dP_{N} \sum_{i =1}^{N} e^{-\beta
E_{\rm c}^{\rm m}} \, \left\{ e^{\beta
E_{\rm c}(x^N_{i-1})} \, \theta(Y_{1}-x^N_i) + e^{\beta E_{\rm c}(x^N_{i})} \,
\theta(x^N_{i-1} -Y_{1}) \right.  \label{eq:22} \\ &\ & + \left.
\left[ e^{\beta E_{\rm c}(x_{i-1}^{N})} \, \theta(x^{N*}_{i-1} - x^{N}_{i}) +
e^{\beta E_{\rm c}(x_{i}^{N})} \, \theta(x^{N}_{i-1} - x^{N*}_{i}) \right]
\theta(x^N_i -Y_{1}) \, \theta(Y_{1}-x^N_{i-1}) \right\} \; , \nonumber
\end{eqnarray}
where $x^{*}$ is that position to the right (left) of $Y_{1}$ where the profile
has the same height as at the point $x$ to the left (right) of $Y_{1}$.  The
average sum over the ballistic intervals contributed by the {\em left-hand}
peak at $X_{0}$ is given by Eq.\ (\ref{eq:20}) with $X=X_{0}$ and
the last term in the brackets omitted, and analogously for the {\em right-hand}
peak at $X_{1}$.  The average sum for the profile of Fig.\ \ref{fig:2} can then
be evaluated as
\begin{equation}
\Xi = 1 + \widetilde{S}/l +
\int_{X_{1}^{*}}^{X_{1}}\frac{dx}{l}e^{-|x-x^{*}|/l}e^{-\beta E_{\rm
c}(X_{0})}  \, \left[ e^{\beta E_{\rm c}(X_{1})}- e^{\beta
E_{\rm c}(x)} \right] \; ,
\label{eq:23}
\end{equation}
where now $ E_{\rm c}^{\rm m} =  E_{\rm c}(X_{0})$ in $\widetilde{S}$.

It is seen from this formula that for $b_{0,1} \ll l \ll w_{1}$, where $w_{1} =
X_{1} - X_{1}^{*}$ is the width of the valley and $b_{0,1}$ are the
``widths'' of the two barriers at $X_{0,1}$, the integral over the
second exponential in the brackets can be neglected. Since
\begin{equation}
\int_{X_{1}^{*}}^{X_{1}} dx \, e^{-|x-x^{*}|/l}  = l (1 - e^{-w_{1}/l})
\approx l \; ,
\label{eq:28aa}
\end{equation}
we obtain
\begin{equation}
\Xi = 1 + \widetilde{S}/l +  e^{-\beta [ E_{\rm c}(X_{0}) -
E_{\rm c}(X_{1})]} \; ,
\label{eq:23aa}
\end{equation}
i.e., we have independent contributions from each peak (first and third terms).

On the other hand, for $S \ll l$ (ballistic regime), the whole integral term in
Eq.\ (\ref{eq:23}) can be neglected (together with the term $\widetilde{S} /l$)
since it is smaller than $(w_{1}/l) \exp\{-\beta [E_{\rm c}(X_{0}) - E_{\rm
c}(X_{1})]\}$.  Then $\Xi$ reduces to the unit term, which represents the
contribution of the transmission across the higher peak at $X_{0}$:  this peak
``eclipses'' the lower peak at $X_{1}$.

\section{Current-voltage characteristic}

\subsection{Classical current-voltage characteristic}

In the foregoing, we have considered single peaks and valleys.  If the profile
contains not just one, but an arbitrary combination of such structures as in a
chain of grains in polycrystalline materials, the average sum $\Xi$ for a
general band edge profile is given by
\begin{equation}
\Xi = 1 + (\widetilde{S} + \widetilde{\Lambda})/l \;
, \label{eq:23a}
\end{equation}
where we have introduced the ``shape term'' $\widetilde{\Lambda}$ for $M$
valleys,
\begin{equation}
\widetilde{\Lambda} = \sum_{v=1}^{M} \int_{X_{v}^{*}}^{X_{v}}dx e^{-|x-x^{*}|/l}
\, e^{-\beta E_{\rm c}^{\rm m}}\left[ e^{\beta E_{\rm c}(X_v)} -
e^{\beta E_{\rm c}(x)}\right] \; .
\label{eq:28}
\end{equation}
The contribution of valley $v$ (with adjoining lower maximum $X_{v}$) consists
of an integral which extends over the width of the valley from $X_{v}^{*}$ to
$X_{v}$.  We now define the effective transport length $L = \Xi \, l $, so that
\begin{equation}
L = l + \widetilde{S} + \widetilde{\Lambda} \; .
\label{eq:44a}
\end{equation}
In compliance with Eq.\ (\ref{eq:28aa}), we find $\widetilde{\Lambda} < S$.

From Eqs.\ (\ref{eq:5}), (\ref{eq:23a}) and (\ref{eq:44a}), we obtain the
classical current-voltage characteristic for nondegenerate systems as
\begin{equation}
j = -ev_{\rm e}N_{\rm c} \, e^{-\beta E_{\rm p}} \, \frac{l}{L} \, (1-
e^{-\beta eV}) \; .
\label{eq:43}
\end{equation}
This formula is the principal result of the present work. It is given here in
the context of classical transport, where its interpretation is most
perspicuous; corrections due to tunneling will be introduced below.

The properties of the current-voltage characteristic (\ref{eq:43}) are
determined by the barrier height $E_{\rm p}$ and the ratio $l/L$.  In
Eq.\ (\ref{eq:44a}) for the effective transport length $L$, the mean
free path $l$ represents the ballistic contribution to the current (this term
is associated with the highest peak of the band edge profile).  The remaining
terms give a quantitative measure of the influence of that part of the electron
motion which is not purely ballistic.  Their contribution amounts to at most
twice the length $S$ of the sample.  The reduced sample length
$\widetilde{S}$ given by Eq.\ (\ref{eq:27aa}) represents a
contribution that characterizes the band edge profile $E_{\rm c}(x)$ in an
integral way; it does not manifestly depend on $l$, only indirectly so via the
profile (cf.\ below).  The shape term $\widetilde{\Lambda}$ given by
Eq.\ (\ref{eq:28}), on the other hand, depends on the detailed structure of
the profile as well as explicitly on the mean free path and thus represents the
interplay of ballistic and diffusive transport.  This term is a distinctive
feature of formula (\ref{eq:43}).

We emphasize that the integrals appearing in the effective transport length
$L$, and in particular the reduced sample length $\widetilde{S}$, result from
averaging over ballistic configurations, {\em not} because $l$ were so small
that the sum over intervals could be replaced with an integral, as assumed in
the diffusive regime.

The barrier height $E_{\rm p}$ and the effective transport length $L$ are
determined by the band edge profile $E_{\rm c}(x)$ which is a solution
of the Poisson equation
\begin{equation}
E_{\rm c}''(x) = \frac{e}{\epsilon_{\rm s}}[ - e n(x) + Q(x)] \; ,
\label{eq:43a}
\end{equation}
where
\begin{equation}
n(x) = N_{\rm c} \, e^{-\beta [E_{\rm c}(x)-E_{\rm F}(x)]}
\label{eq:43ab}
\end{equation}
is the conduction electron density and $Q(x)$ is the density of fixed charges.
In the ``trapping model'' for grain boundaries in polycrystalline materials
\cite{kam71,set75,wei98,wei99}, which our calculations in Sec.\ IV are based
upon, the density  $Q(x)$ is given by
\begin{equation}
Q(x) = e N_{\rm don} + \sum_{v=1}^{M} q^{\rm t}_{v} \, \delta (x-X_{v})
\; ,
\label{eq:43ac}
\end{equation}
where $N_{\rm don}$ is the density of donor atoms (assumed completely ionized),
and $q^{\rm t}_{v}$ is the area density of the charge associated with occupied
acceptor-like ``trapping states'' localized at the grain boundary at $X_{v}$.
The donor density $N_{\rm don}$ not only affects the band edge profile $E_{\rm
c}(x)$, but also determines the magnitude of the mean free path $l$ (a simple
relation between $l$ and $N_{\rm don}$ is obtained from the analytical
expression for the $N_{\rm don}$-dependence of the electron mobility $\mu$
given in Ref.\ \cite{aro82}, using $\mu = ev_{\rm e}\beta l$).  Thus, there is,
in general, an indirect relation between $E_{\rm c}(x)$ and $l$, which implies
an indirect (and generally rather strong) $l$-dependence of the barrier height
$E_{\rm p}$ as well as of the reduced sample length $\widetilde{S}$
and the shape term $\widetilde{\Lambda}$ [as noted above, the latter
term also depends explicitly on $l$, namely via the weight function
$\exp(-|x-x^*|/l)$].

When the bias vanishes, the electron density $n(x)$ has the equilibrium form
given by expression (\ref{eq:43ab}) with $E_{\rm F}(x) = {\rm const.} = E_{\rm
F}(0)$, and we obtain a nonlinear differential equation for $E_{\rm c}(x)$.  In
the presence of bias, the Poisson equation (\ref{eq:43a}) must be solved in
conjunction with an equation for the current $j$.  In the diffusive limit, this
equation is given by the familiar drift-diffusion expression for the current,
\begin{equation}
j = \mu n(x) \frac{d}{dx}E_{\rm F}(x) = \frac{\mu}{\beta} \frac{d}{dx}n(x) +
\mu n(x) \frac{d}{dx}E_{\rm c}(x) \; ,
\label{eq:43c}
\end{equation}
which determines $n(x)$ in terms of $ E_{\rm c}(x)$ and the constant parameter
$j$.  On the other hand, when the current is ballistic in an interval
$(x_{i-1}^{N}, x_{i}^{N})$ [cf.\ Eqs.\ (\ref{eq:1})], only the equilibrium
densities $n(x_{i-1}^{N}), n(x_{i}^{N})$ at the end points of the interval
enter into the description, and an averaging formalism must be provided which
allows one to derive from these discrete values of the density a continuous
physical density to be used in the Poisson equation.

We close this section with a brief discussion of the special case of a single
barrier.  In the case of a single grain boundary or a Schottky contact, the
band edge profile exhibits a single peak and no valleys, so that the shape term
vanishes, $\widetilde{\Lambda} = 0$, and Eqs.\ (\ref{eq:44a}) and (\ref{eq:43})
yield
\begin{equation}
j = - ev_{\rm e}N_{\rm c}\, e^{-\beta E_{p}} \frac{l}{l +
\widetilde{S}} (1-e^{-\beta eV}) \; .
\label{eq:50}
\end{equation}
For a grain boundary, the barrier height $E_{\rm p}$ is given by the difference
of the profile maximum at the boundary and the Fermi level in the bulk of the
grain.  The barrier height of a Schottky contact is equal to the difference of
the profile maximum in the semiconductor and the Fermi level of the metal.
Equation (\ref{eq:50}) is formally identical to Eq.\ (6) of Ref.\ \cite{cro66}
(except for the tunneling correction).  However, the present derivation is
different from that of Ref.\ \cite{cro66}, the averaging over ballistic
configurations being the crucial ingredient.

If the mean free path is much longer than the reduced sample length, $l \gg
\widetilde{S}$, one obtains from Eq.\ (\ref{eq:50}) the
thermionic-emission formula \cite{bet42,sze81}.  In the opposite case, $l \ll
\widetilde{S}$, the transport mechanism is diffusive; with the use of
\begin{equation}
\mu =  e v_{\rm e} \beta l \; ,
\label{eq:51a}
\end{equation}
Eq.\ (\ref{eq:50}) becomes
\begin{equation}
j = - \frac{\mu N_{\rm c}
e^{-\beta E_{p}}}{\beta \widetilde{S}}(1-e^{-\beta eV}) \; .
\label{eq:52}
\end{equation}
Looking at the diffusive limit differentially, we have from Eq.\ (\ref{eq:1})
in conjunction with Eq.\ (\ref{eq:43ab}), replacing the discrete coordinate
$x_{i}^{N}$
with the continuous coordinate $x$, i.e.,\ setting $x_{i}^{N} = il = x$
$(i=1,\ldots , N)$, $l = S/N$ with $ N \rightarrow \infty$,
\begin{equation}
j = \beta ev_{\rm e}N_{\rm c} e^{-\beta [E_{\rm c}(x)-E_{\rm F}(x)]}
[dE_{\rm F}(x)/dx] \cdot l = \mu n(x) \frac{d}{dx} E_{\rm F}(x) \; ,
\label{eq:53}
\end{equation}
in agreement with Eq.\ (\ref{eq:43c}).

\subsection{Quantum effects}

Since we are focussing attention on the case where the mean free path and the
relevant structures of the band edge profile are of comparable length, we are
generally dealing with systems of small dimensions and therefore must expect
quantum effects such as discretization of energy states and tunneling to play a
role.  Considering, e.g., a potential valley with a width of the order of 20
nm, as typified by the examples discussed below, we find that the electron
motion is quantized with energy spacings of about 0.03 eV.  Since the barrier
heights in the examples exceed this value by an order of magnitude, one may
still speak of a classical continuum of states.  On the other hand, the wave
length of the electrons at $T = 300$ K is $\lambda = h/(m^{*}v_{\rm e}) \approx
10$ nm, so that quantum tunneling should be important.  The formalism for
including the effect of tunneling, i.e., for going from thermionic emission to
thermionic field emission, will be developed in the following.  We do not
include the effects of phase interference and localization, since these are not
expected to play an important role for the polycrystalline materials we are
considering.

The transmission probabilities for ballistic intervals with no peaks inside are
treated classically, as before.  Tunneling has to be taken into account near
each peak in intervals $X_{n}^{-} < x < X_{n}^{+}$ containing the peak at
$X_{n}$ $(n=0,1, \ldots ,M)$.  Since we can treat only {\em ballistic} quantum
transport, these intervals must not contain equilibration points.  In other
words, in these intervals we must effectively set $l \rightarrow \infty$, and
in the integrals over $x$ these intervals are omitted, since here $dx/l = 0$.
The lengths to be chosen for the intervals $[X_{n}^{-},X_{n}^{+}]$ will be
discussed below.

In WKB approximation, the thermally averaged quantal probability ${\cal
T}^{\rm WKB}(i)$ for ballistic transmission from $x_{i-1}^{N}$ to $x_{i}^{N}$
is given by
\begin{equation}
{\cal T}^{\rm WKB}(i) = e^{-\beta E_{\rm c}^{\rm m}(i)} + \beta \, \int_{E_{\rm
c}^{(i)}}^{E_{\rm c}^{\rm m}(i)} dE \; \exp\left( -\beta E -\frac{2}{\hbar}
\int_{y_{i}}^{y_{i}^*} dx \{2m^{*}[E_{\rm c}(x) - E] \}^{1/2}
\right) \; ,
\label{eq:21f}
\end{equation}
where $E_{\rm c}^{(i)}$ is defined after Eq.\ (\ref{eq:1aaa}). The limits of
integration $y_{i}$ and $y_{i}^*$ are the turning points at energy $E$ on either side
of the peak (if the interval $\{i \} = [x_{i-1}^{N},x_{i}^{N}]$ contains
several peaks, the integral in the exponential goes from the left-most turning
point to the right-most).

It is found that in the current-voltage characteristic (\ref{eq:43}), the barrier
factor $\exp(- \beta E_{\rm p})$ is to be multiplied by the tunneling
correction ${\cal C}^{-1}$, with
\begin{equation}
{\cal C} = e^{- \beta [E_{\rm c}^{\rm m} - E_{\rm c}(X_{0}^{-})]}
\; \left[ e^{-(X_{M}^{-} - X_{0}^{+})/l} \left(
\frac{1}{{\cal T}^{\rm WKB}(X_{0}^{-},X_{M}^{+})} -  \frac{1}{{\cal T}^{\rm
WKB}(X_{0}^{-},X_{0}^{+})} \right) +  \frac{1}{{\cal T}^{\rm
WKB}(X_{0}^{-},X_{0}^{+})} \right] \; ,
\label{eq:21kk}
\end{equation}
where the notation ${\cal T}^{\rm WKB}(X_{0}^{-},X_{M}^{+})$ refers to the
averaged WKB transmission probability ${\cal T}^{\rm WKB}(i)$ for the interval
$[X_{0}^{-},X_{M}^{+}]$, etc; here, $X_0$ is the position of the overall maximum peak
$E_{\rm c}^{\rm m}$.  The reduced sample length $\widetilde{S}$ and
the shape term $\widetilde{\Lambda}$ that enter the effective transport length
$L$ [cf.\ Eq.\ (\ref{eq:44a})] now appear as
\begin{equation}
\widetilde{S} = {\cal C}^{-1} \, \int_{0}^{S^{\bullet}} dx \, e^{- \beta
[E_{\rm c}^{\rm m} - E_{\rm c}(x)]} \; ,
 \label{eq:21fff}
\end{equation}
where the upper limit $S^{\bullet}$ implies that all intervals
$[X_{n}^{-},X_{n}^{+}]$ $(n = 0,1, \ldots, M)$ are to be omitted,
and
\begin{equation}
\widetilde{\Lambda} = {\cal C}^{-1}\sum _{v=1}^{M} \; \int_{X_{v}^{-
\bullet}}^{X_{v}^{-}} dx \, e^{-|x-x^{*}|/l} \, e^{-\beta E_{\rm c}^{\rm m}}
\left[ A_{v} \, e^{\beta E_{\rm c}(X_{v})} - e^{\beta E_{\rm c}(x)} \right] \;
\; ,
\label{eq:21ffg}
\end{equation}
with $X_{v}^{- \bullet} = \max\{X_{v}^{- *},X_{v-1}^{+}\}$ and
\begin{equation}
A_{v} = {\cal T}(X_{v}^{-},X_{v}^{+})/{\cal T}^{\rm
WKB}(X_{v}^{-},X_{v}^{+}) \; , \hspace*{0.5cm} v=1,\ldots,M \; ,
\label{eq:21k}
\end{equation}
where ${\cal T}$ is the average classical transmission probability
(\ref{eq:1aaa}).  It can be shown that $\widetilde{\Lambda} < S$, as in the
classical case.

For a single barrier, Eq.\ (\ref{eq:50}) is generalized to
\begin{equation}
j = - ev_{\rm e}N_{\rm c}\, {\cal C}^{-1} \, e^{-\beta E_{p}} \frac{l}{l +
\widetilde{S}} (1-e^{-\beta eV}) \; .
\label{eq:50aaa}
\end{equation}
According to Eq.\ (\ref{eq:21fff}), $\widetilde{S}$ includes the tunneling
correction factor ${\cal C}^{-1} \geq 1$, and thus tunneling enhances the
relative effect of the non-ballistic part of the electron motion embodied in
the term $\widetilde{S}$ in the denominator of expression (\ref{eq:50aaa}).

The intervals $[X_{n}^{-},X_{n}^{+}]$ enclosing the peaks at $X_{n}$ are
defined as the intervals in which tunneling plays a role; they are determined
by the requirement that the ratio $A_{v}$ becomes virtually independent of the
interval if that extends beyond the end points $X_{n}^{\pm}$.  On the other
hand, it must be ascertained that these lengths are smaller than the mean free
path $l$, otherwise the quantum correction scheme breaks down:  if tunneling
takes place over distances larger than the mean free path, thermalization and
tunneling occur simultaneously, which cannot be described in the present
framework.  Explicit calculations of the quantum corrections will be carried
out below.

\subsection{The degenerate case}

In the degenerate case, a simple treatment is possible only in the limit of
zero bias. Formula (\ref{eq:1aa}) for the current is to be replaced (for
infinitesimal bias) with
\begin{equation}
j(i) = \frac{4\pi em^{*}}{\beta h^{3}} \; \delta_i \;
\frac{\partial}{\partial E_{\rm F}} \int_{E_{\rm c}^{(i)}}^{\infty}
dE \, T_i(E) \,  \ln (1+ e^{-\beta [E - E_{\rm F}]}) \; ,
\label{eq:62}
\end{equation}
where we have set $E_{\rm F}(x_{i-1}^{N}) = E_{\rm F}$ and $E_{\rm
F}(x_{i}^{N}) = E_{\rm F} + \delta_i$.  Classically, this becomes
\begin{equation}
j(i) = \frac{4\pi em^{*}}{\beta h^{3}} \delta_i \;
\ln (1+ e^{-\beta[E_{\rm c}^{\rm m}(i)- E_{\rm F}] })\; .
\label{eq:62a}
\end{equation}
By analogy with Eq.\ (\ref{eq:3}), we then find
\begin{equation}
E_{\rm F}(S) - E_{\rm F}(0) = \sum_{i=1}^{N} \delta_i =
j \, \frac{\beta h^{3}}{4\pi e m^{*}}
\sum_{i=1}^{N}\frac{1}{\ln (1+ e^{\beta [E_{\rm F} -
E_{\rm c}^{\rm m}(i)]})} \; ,
\label{eq:65}
\end{equation}
with $E_{\rm F}(0)$ and $E_{\rm F}(S)$ differing infinitesimally from the
equilibrium value $E_{\rm F}$. Denoting here the average over the
sum (cf.\ Sec.\ II.A), multiplied by $\ln (1+ e^{\beta [E_{\rm F} - E_{\rm
c}^{\rm m}]})$, by $\Xi_{\rm d}$,
\begin{equation}
\Xi_{\rm d} = \left\langle \sum_{i=1}^{N}\frac{\ln (1+ e^{\beta [E_{\rm
F} - E_{\rm c}^{\rm m}]})}{\ln (1+ e^{\beta [E_{\rm
F} - E_{\rm c}^{\rm m}(i)]})}\right\rangle_{\{N, x^N_{i} \}}  \; ,
\label{eq:66}
\end{equation}
we have
\begin{equation}
eV = E_{\rm F}(S) - E_{\rm F}(0) = - j \; \frac{\beta h^{3}}{4 \pi e m^{*}} \; \frac{1}{\ln (1+ e^{\beta [E_{\rm
F} - E_{\rm c}^{\rm m}]})} \; \Xi_{\rm d} \; .
\label{eq:65a}
\end{equation}
The calculation of the average sum $\Xi_{\rm d}$ follows Sec.\ II.B.
A (classical) effective transport length $L_{\rm d}$ is introduced as
\begin{equation}
L_{\rm d} = \Xi_{\rm d} \, l = l + \widetilde{S}_{\rm d}
+ \widetilde{\Lambda}_{\rm d} \; , \label{eq:66b}
\end{equation}
where
\begin{equation}
\widetilde{S}_{\rm d} = \int_{0}^{S} dx \; \frac{\ln (1+ e^{\beta [E_{\rm F} -
E_{\rm c}^{\rm m }]})}{\ln (1+ e^{\beta [E_{\rm F} - E_{\rm c}(x)]})}
\label{eq:68}
\end{equation}
is the reduced sample length for the degenerate case, in
generalization of Eq.\ (\ref{eq:27aa}), and
\begin{equation}
\widetilde{\Lambda}_{\rm d} = \sum_{v=1}^{M} \int_{X_{v}^{*}}^{X_{v}} dx \,
e^{-|x-x^{*}_{v}|/l} \left[ \frac{\ln (1+ e^{\beta [E_{\rm F} - E_{\rm c}^{\rm
m }]})}{\ln (1+ e^{\beta [E_{\rm F} - E_{\rm c}(X_{v})]})} - \frac {\ln (1+
e^{\beta [E_{\rm F} - E_{\rm c}^{\rm m }]})}{\ln (1+ e^{\beta [E_{\rm F} -
E_{\rm c}(x)]})} \right]
\label{eq:69}
\end{equation}
is the generalization of the shape term (\ref{eq:28}). The zero-bias
conductance per unit area, finally, is obtained as
\begin{equation}
g = \left( \frac{|j|}{V} \right)_{V \rightarrow 0} = \frac{4\pi e^{2} m^{*}
}{\beta h^{3}} \;  \ln (1+ e^{-\beta E_{\rm p}}) \; \frac{l}{L_{\rm d}} \;  ,
\label{eq:70}
\end{equation}
with $E_{\rm p} = E_{\rm c}^{\rm m} - E_{\rm F}$ (note that in the degenerate
case, $E_{\rm p}$ may be negative). Quantum effects may be taken into account
as in the nondegenerate case.

\subsection{The effect of dimensionality}

The samples we consider are assumed to have a parallel-plane structure and thus
have an essentially one-dimensional geometry.  Nevertheless, the treatment of
the transport process itself should, of course, be three-dimensional.
Therefore, in the discussion preceding Eq.\ (\ref{eq:6}), one must consider the
distance between collisions, $\xi$, in three-dimensional space. This introduces
a polar angle to be summed over appropriately in averaging over the ballistic
configurations, thereby rendering the formalism appreciably more complicated.
In our effort to obtain a simple, {\em closed} formula for the current-voltage
characteristic, we have restricted ourselves to a one-dimensional formulation,
which leads to the desired result in a straightforward way.

Now, for the special case of a {\em constant} profile, $E_{\rm c}(x) = 0$, de
Jong \cite{dej94} has derived a formula which describes the transition from the
(ballistic) Sharvin resistance \cite{sha65} to the (diffusive) Drude zero-bias
conductance \cite{dru00,max91}.  In this work, use is made of an
integro-differential equation in two and three dimensions, leading to a
numerical result which, for the three-dimensional case, is summarized in the
formula (in the notation of our Eq.\ (\ref{eq:70}) with $E_{\rm c}^{\rm m} = 0$
and $\beta E_{\rm F} \gg 1$)

\begin{equation}
g = \frac{4\pi e^{2} m^{*} E_{\rm F}}{h^{3}} \; \frac{l}{l + \smfrac{3}{4}
\gamma S} \;
, \label{eq:70aaa}
\end{equation}
where $\gamma$ increases monotonically from 1 to $4/3$ as $l/S$ goes from 0
(diffusive limit) to $\infty$ (ballistic limit). In our one-dimensional
formulation, the denominator in formula (\ref{eq:70aaa}) reads simply $l+S$. We
therefore suggest that the restriction to a one-dimensional formulation
introduces a numerical error of perhaps 30 percent; the main features of our
model, which the following examples show to involve enhancements by
factors of 10, should not be severely affected.

Unfortunately, one cannot make use of differential equations in the more
general case when the profile $E_{\rm c}(x)$ changes significantly over
distances of the order of the mean free path $l$. Anyhow, we are not so much
interested in the numerical solution of the transport problem (which would most
efficiently be handled by Monte Carlo simulations) but, as far as possible, in
its physical analysis via simple and transparent closed formulas.

\section{Examples}

\subsection{Zero-bias conductivity of a chain of identical grains}

The zero-bias conductance per unit area, $g = (|j|/V)_{V \rightarrow 0}$, is
obtained as
\begin{equation}
g = \beta e^{2}v_{\rm e}N_{\rm c} \, {\cal C}^{-1} \, e^{-\beta E_{\rm p}} \;
\frac{l}{L} = \frac{4\pi e^{2} m^{*}}{\beta h^{3}} \, {\cal C}^{-1} \,
e^{-\beta E_{\rm p}} \; \frac{l}{L} \; .
\label{eq:56}
\end{equation}
The band edge profile $E_{\rm c}(x)$ for zero bias is given by its equilibrium
shape, and is calculated as a solution of the Poisson equation (\ref{eq:43a})
in the trapping model.

\begin{figure}
\epsfysize=4.5cm
\epsfbox[7 653 266 743]{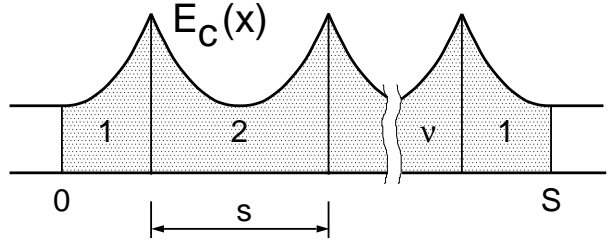}
\caption{
Equilibrium band edge profile $E_{\rm c}(x)$ of a chain of $\nu$ identical
grains of length $s$, forming a sample of total length $S = \nu s$.
}
\vspace{0.4cm}
\label{fig:3}
\end{figure}

We consider a chain of $\nu$ identical grains, each of length $s$, as shown in
Fig.\ \ref{fig:3}.  There are $\nu-1$ identical valleys in $E_{\rm c}(x)$.
Writing $\widetilde{S} = \nu \widetilde{s}$ and $\widetilde{\Lambda} = (\nu -1)
\widetilde{l}$, we have from Eq.\ (\ref{eq:44a})
\begin{equation}
L = l + \nu \widetilde{s} + (\nu-1) \widetilde{l} \; .
\label{eq:58}
\end{equation}
As mentioned above, the profile $E_{\rm c}(x)$, and therefore also the barrier
height $E_{\rm p}$, the reduced grain length $\widetilde{s}$, and the
single-grain shape term $\widetilde{l}$, indirectly depend on the mean free
path $l$ via its connection with the donor density $N_{\rm don}$.  The
$l$-dependence of the main feature of the profile, the barrier height $E_{\rm
p}$, can be read from the $l$-dependence of the conductance $g$ for a single
grain in the ballistic regime, when $L=l$ in Eq.\ (\ref{eq:56}).

The term $l$ in expression (\ref{eq:58}) for the effective transport length $L$
represents the ballistic contribution to the current.  For purely ballistic
motion ($l \gg \nu s$ and $L \rightarrow l$), one finds from Eq.\ (\ref{eq:56})
in the classical limit (${\cal C}=1$) that only {\em one} barrier, the front
barrier, is relevant since it eclipses all others (cf.\ above, and also the
discussion in Sec.\ II of Ref.\ \cite{kim84}).  However, when tunneling takes
place (possibly through several neighboring peaks), Eq.\ (\ref{eq:50aaa}) in
conjunction with Eq.\ (\ref{eq:21kk}) is to be used to obtain the conductance.

\begin{figure}
\epsfysize=10.0cm
\epsfbox [-100 493 251 780]{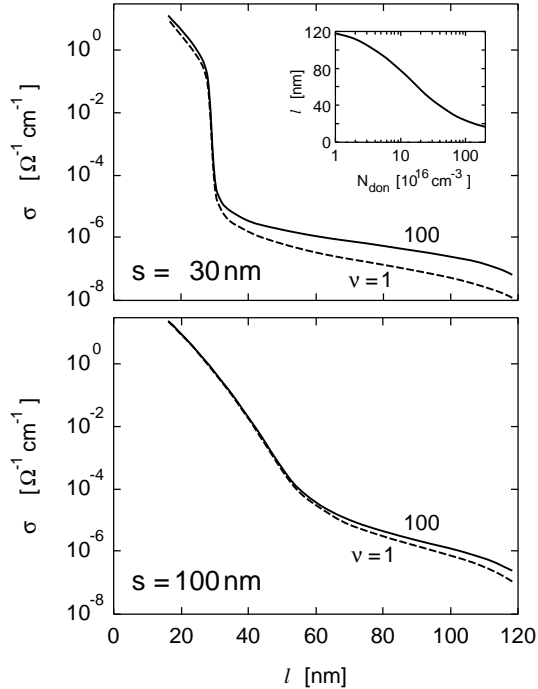}
\caption{
Zero-bias conductivity $\sigma$ in the unified model for Si samples with grain
length $s$ = 30 nm (upper panel) and grain length $s$ = 100 nm (lower panel),
plotted as a function of mean free path $l$ for temperature $T=300$ K, trapping
state density $N_{\rm t} = 2 \times 10^{12}$ cm$^{-2}$, and a single trapping
level located at 0.56 eV above the valence band edge.  Dashed curves:  single
grain ($\nu =1$); solid curves:  chain of a hundred grains ($\nu = 100$).  The
inset in the upper panel shows the relation between $l$ and the donor density
$N_{\rm don}$ at $T = 300 $ K.
}
\vspace{0.4cm}
\label{fig:4}
\end{figure}

For the purpose of illustration, we consider two silicon samples with grain
length $s = 30$ nm ($\mu$c-Si) and $s= 100$ nm (pc-Si), respectively. Instead
of the zero-bias conductance $g$, we introduce here the zero-bias conductivity
$\sigma$ as conductance times sample length,
\begin{equation}
\sigma = g S = g \nu s
\label{eq:59}
\end{equation}
(in the diffusive limit, $\sigma$ is independent of the number of grains
$\nu$).  The conductivity has been calculated at temperature $T=300$ K by means
of formulas (\ref{eq:56}) and (\ref{eq:58}) as a function of the donor density
$N_{\rm don}$.  Applying the criteria discussed in the final paragraph of Sec.\
III.B, the length of the ``tunneling intervals'' $[X_{n}^{-},X_{n}^{+}]$
enclosing the peaks was taken to be 6 nm (which, not surprisingly, is of the
order of the wave length $\lambda = 10$ nm of the electrons); this value is
much smaller than the magnitude of the mean free path in all cases considered.
The results are shown in Fig.\ \ref{fig:4} as a function of the {\em mean free
path} $l$, employing the connection between $l$ and $N_{\rm don}$ obtained from
Ref.\ \cite{aro82} and displayed in the inset.  It is seen that the
conductivity depends appreciably on the number of grains in the case of the
smaller grain size, $s=30$ nm. Thus it emerges that the conductivity
calculated for a single grain cannot simply be carried over to the entire
microcrystalline sample.  The latter must be considered as a whole; there is in
this case no grain-specific conductivity.

Tunneling has a direct influence on the barrier transmission probability, which
is taken into account through the correction factor ${\cal C}^{-1}$ in Eq.\
(\ref{eq:56}). It enhances the transmission probability by up to 50\% in the
region of small $l$.

\begin{figure}
\epsfysize=10.0cm
\epsfbox[-108 473 438 788]{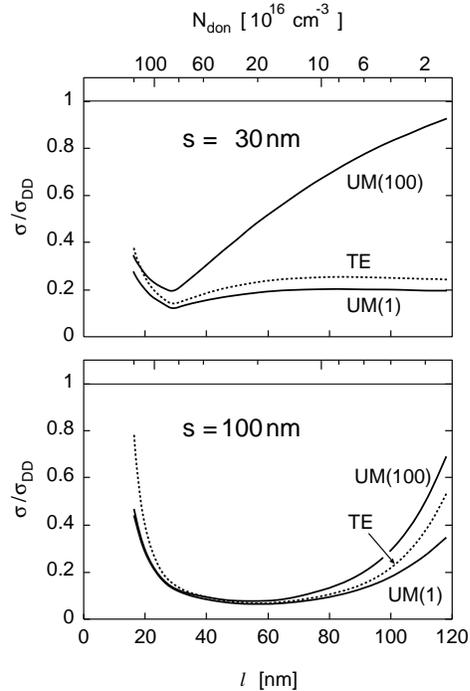}
\caption{
Relative conductivities $\sigma/\sigma_{\rm DD}$ for the cases of Fig.\
\ref{fig:4}.
}
\vspace{0.4cm}
\label{fig:5}
\end{figure}

In Fig.\ \ref{fig:5}, we summarize the results of the unified model in
comparison to those of the drift-diffusion and thermionic-emission models.
Since tunneling does not affect appreciably the mutual relation between the
different transport models, we here consider only the {\em classical}
conductivities, and plot these {\em relative to the conductivity within the
drift-diffusion model, $\sigma_{\rm DD}$} [with $L = \nu \widetilde{s}$ in Eq.\
(\ref{eq:56})].  The curves labelled UM(1) and UM(100) represent the
conductivity of the unified model for a chain of one and a hundred grains,
respectively, divided by the conductivity calculated within the drift-diffusion
model.  In the regions where these curves approach unity, the transport
mechanism is predominantly diffusive.  The curve labelled TE (``thermionic
emission'') represents the relative conductivity for purely ballistic transport
across a {\em single} grain boundary, the result being identified (as is
usually done) with the conductivity of the entire chain.  This procedure
ignores the eclipsing effect alluded to above, and is justified only if the
mean free path is long compared to the width of the barrier but short compared
to the length of the grain [cf.\ the conditions leading to Eq.\
(\ref{eq:23aa})], so that while moving through the grain, the electrons are
thermalized and ``face'' each grain boundary barrier with the same thermal
distribution as when passing over the previous one.

We see that the transport mechanism tends to become diffusive for very small
values of $l$, and also for $l > 80$ nm in the case $\nu = 100$; this holds for
$s = 30$ nm as well as for $s = 100$ nm.  The thermionic-emission model yields
acceptable results in the region $20 < l < 60$ nm at $s=100$ nm for $\nu = 1$
and $\nu = 100$.  At $s=30$ nm, it is valid for $\nu = 1$ but completely
invalid for $\nu = 100$.  In the latter case, neither the drift-diffusion nor
the thermionic emission models describe adequately the correct mechanism [which
is represented by the curve labelled UM(100)].

The transport properties have been discussed here for just two grain lengths
$s$ representative of microcrystalline and polycrystalline silicon,
respectively.  A more comprehensive study would have to deal with the
$s$-dependence over a suitably broad range, which in particular may lead to the
disclosure of possible scaling properties.

\subsection{Degenerate case: Single barrier}

We apply formula (\ref{eq:70}) to transport through a single grain boundary.
Choosing parameter values so as to reproduce approximately the conditions of
the example of Ref.\ \cite{pri98}, we consider the temperature dependence of
the conductance for a grain boundary barrier in highly doped pc-SnO$_{2}$:Sb
with grain length $s = 50$ nm.  We have calculated the equilibrium band edge
profile $E_{\rm c}(x)$ at $T=300$ K and $N_{\rm don} = 7.2 \times 10^{18}$
cm$^{-3}$, corresponding to a mean free path $l = 11$ nm.  We assume a single
trapping level at midgap (1.75 eV above the valance band edge) with trapping
state density $N_{\rm t} = 1.95 \times 10^{12}$ cm$^{-2}$.  The barrier height
is obtained as $E_{\rm p} = 0.038$ eV and the barrier width as $\approx 6$ nm.
Since we intend to apply formula (\ref{eq:70}) in a schematic manner only and
the band edge profile is found to depend only weakly on $T$ and $N_{\rm don}$
(or $E_{\rm F}$), we adopt the profile calculated at $T=300$ K for all values
of $T$ and control the degree of degeneracy by independently choosing the value
of $E_{\rm F}$ in Eq.\ (\ref{eq:70}).

\begin{figure}
\epsfysize=6.0cm
\epsfbox[-125 588 251 786]{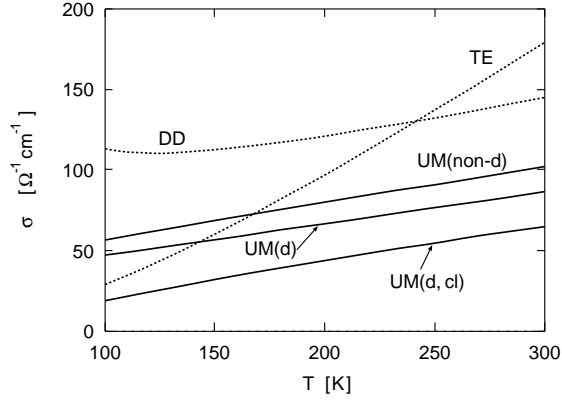}
\caption{
Temperature dependence of the barrier conductivity $\sigma$ for a grain
boundary barrier in pc-SnO$_{2}$:Sb with conduction band edge profile
calculated for the fixed parameter values $s = 50$ nm, $T = 300$ K, $N_{\rm
don} = 7.2 \times 10^{18}$ cm$^{-3}$ ($l = 11$ nm), $N_{\rm t} = 1.95 \times
10^{12}$ cm$^{-2}$, and a single trapping level located at 1.75 eV above the
valence band edge.  The Fermi level is chosen as $E_{\rm F} = 3.52$ eV.  UM(d):
unified model for the degenerate case (Fermi-Dirac statistics); UM(non-d):
unified model for the nondegenerate case (Boltzmann statistics); DD:
drift-diffusion model; TE:  ballistic model;  UM(d, cl): unified model for the
degenerate classical case (no tunneling).
}
\vspace{0.4cm}
\label{fig:6}
\end{figure}

In Fig.\ \ref{fig:6}, we display the $T$-dependence of the barrier
conductivity $\sigma = gS$, calculated from Eq.\ (\ref{eq:70}) using for the
mean free path the fixed value $l = 11$ nm (when comparing with the results of
Ref.\ \cite{pri98}, one must take account of the factor $\alpha_{\rm eff}
\approx 10^{-2}$ introduced there).  The case considered here [$E_{\rm F} =
3.52$ eV, compared to $E_{\rm c}$(grain bulk) = 3.50 eV] is strongly
degenerate; nevertheless, the results of calculations using Boltzmann and
Fermi-Dirac statistics [UM(non-d) vs.\ UM(d)] are not too far apart.  Further,
it is observed, by comparing the curves labelled UM(d) and UM(d,cl), that the
effect of tunneling increases the conductivity by more than a factor of 2,
becoming stronger as the temperature decreases.

The purely diffusive and ballistic (degenerate) conductivities are also shown
in Fig.\ \ref{fig:6}.  It appears that, as the temperature rises, the
transport character in the unified model changes from ballistic to diffusive.
This is governed by the relative size of the terms $l$ and $\widetilde{S}_{\rm
d}$ in the effective transport length $L_{\rm d}$ appearing in the denominator
of expression (\ref{eq:70}) for the conductance $g$.

Formula (\ref{eq:68}) yields for $\widetilde{S}_{\rm d}$ a value which
increases with $E_{\rm F}$, i.e., with donor density $N_{\rm don}$.  In Ref.\
\cite{pri98}, the corresponding term $\smfrac{3}{4} w$ is found (by fitting to
data) to decrease instead; a more detailed analysis appears to be necessary for
an explanation of this discrepancy.

\section{Summary and conclusions}

For electron transport in parallel-plane semiconducting structures, we have
developed a generalized Drude model which unifies ballistic and diffusive
transport for arbitrary magnitude of the mean free path and arbitrary shape of
the conduction band edge profile.  The semiclassical approach has been adopted,
but tunneling has been taken into account in WKB approximation.

The basic assumption of the model is that the electrons move ballistically over
intervals whose lengths are randomly distributed about the value of the mean
free path.  By averaging over the random configurations of ballistic intervals,
we have derived simple formulas for the current-voltage characteristic (in the
nondegenerate case) and for the zero-bias conductance (in the degenerate case).
The distinctive feature of these formulas is the presence of an effective
length that comprises a shape term directly manifesting the interplay of
ballistic and diffusive transport.  Previously obtained formulas for the
current-voltage characteristic and for the zero-bias conductance refer to
special cases and do not include such a term.

We have performed numerical calculations of the zero-bias conductivity for
chains of grains of (nondegenerate) $\mu$c-Si and pc-Si, and for a single grain
boundary in highly doped (degenerate) pc-SnO$_{2}$:Sb.  The calculations for Si
show substantial deviations of the results of the unified model from those of
the purely ballistic and purely diffusive models.  Moreover, within the unified
model, one finds a fairly strong dependence of the conductivity on the number
of grains.

For the calculation of the zero-bias conductivity, the band edge profile to be
used is the equilibrium profile.  Except in this case and in the diffusive
limit, the determination of the band edge profile is highly complicated in
general.  Basically, one has to solve the Poisson equation along with the
relevant current equation self-consistently, taking into account the averaging
of the electron density over configurations of ballistic intervals.

In this work, we have considered transport in one dimension; it has been argued
that this should be sufficient to exhibit the essential features of the
physical phenomena involved.  Possible generalizations would be the inclusion
of the coupling to minority carriers, of optically induced carrier generation,
and of recombination.

{}

\end{document}